\input amstex
\documentstyle{amsppt}
\topmatter
\title
Causal Monotonicity, Omniscient Foliations, and the Shape of
Space\endtitle
\author
Steven G. Harris \\ Robert J. Low\endauthor
\address
Department of Mathematics, Saint Louis University, St.
Louis, MO 63103,  USA\endaddress
\email
harrissg\@slu.edu\endemail
\address
Department of Mathematics, Coventry University, Coventry CV1
5FB, UK\endaddress
\email
r.low\@coventry.ac.uk\endemail
\thanks
Support from the Aspen center for Physics is gratefully
acknowledged.\endthanks

\abstract
What is the shape of space in a spacetime? One way of
addressing this issue is to consider edgeless spacelike
submanifolds of the spacetime. An alternative is to
foliate the spacetime by timelike curves and consider 
the quotient obtained by identifying points on the same 
timelike curve. In this article we investigate each of these
notions and obtain conditions such that it yields a meaningful 
shape of space. We also consider the relationship between 
these two notions and find conditions for the quotient 
space to be diffeomorphic to any edgeless spacelike 
hypersurface.  In particular, we find conditions in which 
merely local behavior (being spacelike) combined with the
correct behavior on the homotopy level guarantees that a
putative shape of space really is precisely that.
\endabstract

\endtopmatter

\define\F{\Cal F}
\define\g{\gamma}
\document

\head
0. Introduction \endhead

What is the shape of space?

There are two questions comprised in that query:  First, what
is, or  ought to be, meant by the phrase, ``shape of space"? 
Second, assuming  we know what this means, what shape does space
have?

By ``shape of space", we will mean (roughly) the diffeomorphism
class of  any (sufficiently well-behaved) edgeless spacelike
hypersurfaces in a  spacetime, assuming there is only one such
class; this clearly begs the  question of which spacetimes are
so favored as to have a shape of space in this
sense.   Part of the purpose of this paper is to give a good
indication of a large class of spacetimes having a well-defined
shape of space, namely,  those with a distinguished class of
timelike curves foliating the  spacetime, obeying some mild
restrictions.  Given such a spacetime, we  can address ourselves
to the second question:  The shape of space is  found to be the
leaf space of the foliation.

It should be noted that everything in this paper is conformally 
invariant:  This is an inquiry into the global causal structure
of spacetimes.

It needs to be emphasised that the shape of space considered
here is generally quite different from the concept of a Cauchy
surface.  In part this is because the spacetimes considered in
this paper are not restricted to being globally hyperbolic,
which is a prerequisite for having a Cauchy surface. 
Furthermore, a Cauchy surface need not be a shape of space: 
the de Sitter spacetime has Cauchy surfaces which are spheres
but also has perfectly well-behaved edgeless spacelike
hypersurfaces diffeomorphic to $\Bbb R^3$.  A rough and ready
rule:  A spacetime with a timelike
boundary cannot have a Cauchy surface (but may have a shape of
space); a spacetime with a spacelike boundary cannot have a
shape of space (but may have a Cauchy surface).  It is only in a
spacetime with only null boundaries that the two might coexist
(and then must be homeomorphic). 

Section 1 explores some general results for foliations of
spacetimes by  timelike curves; it turns out that the crucial
matter is whether or  not the leaf space is Hausdorff. 
Section 2 explores how this is related  to the concept of causal
monotonicity for a complete timelike vector  field tangent to
the foliation.  Section 3 makes use of the notion of an
omniscient foliation and specializes to the case of 
conformastationary spacetimes.  Section 4 treats an application
to  strongly causal spacetimes, giving conditions under which
all locally ``nice" spacelike hypersurfaces are globally
well-behaved (and so have the proper shape of space).

Several of the results in this paper were announced, in
preliminary form, in \cite{HL}.

\head
1. Timelike Foliations \endhead

Throughout this paper, $V$ will denote a spacetime (a
time-oriented  Lorentz manifold: paracompact and Hausdorff),
$\F$ a differentiable  foliation of V by timelike curves, and
$Q$ the leaf space of $\F$, i.e.,  the quotient space (with
quotient topology) of $V$ modulo the  equivalence relation that
two points are related iff they lie on the  same leaf (i.e.,
curve) in $\F$; in short, $Q = V/\F$.

A typical example of a timelike foliation on a spacetime $V$ is
the set  of integral curves of some timelike vector field $U$. 
In particular, if  $U$ is complete (i.e., each integral curve
$\g$ is defined as a map from  all of $\Bbb R$ into $V$), then
there is an induced group action of  $\Bbb R$ on $V$:  For any
$x \in V$, let $\g_x$ be the integral curve of $U$ with
$\g_x(0) = x$; then for $t \in \Bbb R$, $t\cdot x = \g_x(t)$.
This is a convenient way to have things arranged because then
$\F$  consists of the orbits of the group action, and $Q = V/\Bbb
R$, the  orbit space.  In fact, so long as $V$ is chronological
(no closed  timelike loops), any timelike foliation can be
expressed in this manner:

\proclaim{Lemma 1.1}
If $\F$ is a timelike foliation in a chronological spacetime
$V$, then  there is a complete vector field whose integral
curves are $\F$.\endproclaim

\demo{Proof}
First observe that since $V$ is chronological, the leaves of
$\F$ are  all diffeomorphic to the line, not the circle.  Next,
observe that any  such foliate $\g$ must exit every compact set
$K$:  Otherwise, there  must be some point $p$ in $K$ which is
not on $\g$ but is an  accumulation point of $\g$.  Consider a
``flow box" around $p$: a  (small) embedded hypersurface $P$
through $p$, transverse to $\F$, and  for each $x$ in $P$ a
piece $\g_x$ of the foliate through $x$, such that  $\bigcup
\{\g_x \;|\; x \in P\}$ is a neighborhood of $p$.  Note that for
all $x$ in $P$ sufficiently close to $p$, $\g_x$ must intersect
both the future and  the past of $p$ ($\g_p$ surely does so, and
$\g_x$ is close to $\g_p$ for $x$ close to $p$).  We have that
for some sequence of points  $\{x_n\}$ in $P$ approaching $p$,
$\g^n = \g_{x_n}$ is actually a part  of $\g$ (so that $p$ can be
approached by $\g$).  Restricting attention  to $n$ being
sufficiently large, we can find $q_n$ and $r_n$ as points  on
$\g^n$ respectively to the future and past of $p$.  We have for
all such  $n$ and $m$, all of $\g^n$ is to the future of  all of
$\g^m$ or {\it  vice versa}, depending on how the segments are
situated on $\g$.   Assuming the former, then we have $p \gg r_n
\gg q_m \gg p$, violating the  chronology condition.

Since $V$ is time-oriented, $\F$ is orientable; therefore, we
can pick a  differentiable vector field $U$ everywhere tangent
to $\F$.  We will  construct a positive scalar function $\lambda
: V \to \Bbb R$ such that  $\lambda U$ is complete.

Each foliate has a parametrization making it an integral curve of
$U$; there is ambiguity in this parametrization (up to an
additive constant), but  we don't care whether or not it can be
done in a globally continuous  manner.  Note that $\lambda U$ will
be complete iff for every foliate $\g: (a,b) \to \Bbb R$, 
$\int_t^b (1/\lambda)\circ\g = \infty$ and
$\int_a^t (1/\lambda)\circ \g = \infty$, where $t$ is any
number in the domain.  Now let
$\{K_n\}$ be an exhaustion of $V$ by compact sets:  $K_n$ is a
subset of the interior of $K_{n+1}$ and $\bigcup_{n=1}^\infty K_n
= V$ (since $V$ is chronological, it is not compact, so this can
be done with an infinite family of compact sets).   For each
$n$, there is some $\delta_n > 0$ such that for  every foliate
$\g$ intersecting $K_n$, the parameter for $\g$ increases  by at
least $\delta_n$ as $\g$ extends from where it exits $K_n$ for
the  last time to where it exits $K_{n+1}$ for the first time. 
Now let  $\lambda$ be any positive scalar function on $V$ such
that on $K_{n+1}-K_n$,  $\lambda < \delta_n$.  Then over each of
the intervals mentioned above,  $\int (1/\lambda)\circ\g >1$;
therefore, $\int_t^b (1/\lambda)\circ\g =  \infty$.  Similarly, we
can arrange it so that $\int_a^t  (1/\lambda)\circ\g =
\infty$.  \qed\enddemo

From now on, we will take $\F$ to be given as the orbit space of
an  $\Bbb R$-action, with the corresponding vector field $U$
being future-directed.  Note that this makes $V$ the total
space of a principal line  bundle over $Q$.  Now, any line
bundle (with a real action) over a true manifold (Hausdorff and
paracompact) admits a global cross-section (see \cite{KN},
Theorem  I.5.7), so the total space is diffeomorphic to
$\Bbb R\times$(base  space) (see, e.g.,
\cite{H1, Corollary 1}).  In \cite{H1, Theorem 2} it  was shown
$Q$ must be a {\it near manifold\/}: a topological space with  a
differentiable atlas and a countable basis; only Hausdorffness
is  lacking in order for it to be a true manifold
(paracompactness would  follow from Hausdorffness).  

Here's a typical example with a non-Hausdorff leaf space:  With
$\Bbb  L^n$ denoting Minkowski $n$-space, with
orthogonal coordinates $\{t,x^1,...,x^{n-1}\}$,  let $V = \Bbb L^3
- \{(0,0,0)\}$.  For each $(x,y) \neq (0,0) = \bold 0$, let
$L_{x,y}$ be the line  $\{x^1 = x, x^2 = y\}$; let
$L_\bold 0^+$ and $L_\bold 0^-$, respectively, denote the $t>0$
and $t<0$ half-lines making up $\{x^1 = 0, x^2 =  0\}$.  Then let
$\F = \{L_{x,y} \;|\; (x,y) \neq  (0,0)\} \cup
\{L_\bold 0^+,L_\bold 0^-\}$.  The points of the leaf space
$Q$ can clearly be labeled by $\{(x,y) \;|\; (x,y) \neq  \bold 0\}
\cup \{\bold 0^+,\bold 0^-\}$.  The neighborhoods of each 
$(x,y)$ are given by the ``tubular" (i.e., $\Bbb R$-invariant) 
neighborhoods of $L_{x,y}$; these are just reflections of the 
neighborhoods of $(x,y)$ in the plane, $\Bbb R^2$.  However, the
tubular neighborhoods of $L_\bold 0^+$ and of $L_\bold 0^-$ all
intersect one  another (alternatively stated: any set of foliates
forming a  neighborhood of a point on $L_\bold 0^+$ must overlap
with any set of foliates doing similar duty for $L_\bold
0^-$); it follows that, in the  quotient topology on $Q$, all
neighborhoods of $\bold 0^+$ and of $\bold  0^-$ intersect with
one another: $\{\bold 0^+,\bold 0^-\}$ form a {\it  non-Hausdorff
pair\/}.  Thus, $Q$ is a classic near-manifold:  $\Bbb  R^2$ with
the origin replaced by a double-point.  Note that, in 
particular, $V \neq \Bbb R \times Q$, since $V$ is Hausdorff,
while $Q$  is not.

There is a particular feature of the causal structure of $\F$ in
the  example above that is worth calling attention to:  The
foliates  corresponding to the non-Hausdorff pair in $Q$, i.e.,
$L_\bold 0^+$ and  $L_\bold 0^-$, have the property that any
point in the first is to the  future of any point in the second;
in general, when this happens, we  will say that the second
foliate is {\it ancestral\/} to the first, and we will call the
two foliates an {\it ancestral pair\/}.   In  fact, this is the
key to discovering that a leaf space is Hausdorff (announced
as Theorem 1 of \cite{HL}):

\proclaim{Theorem 1.2}
Let $V$ be a chronological spacetime and let $\F$ be a
foliation of $V$  by timelike curves.  If $\F$ contains no
ancestral pairs, then the leaf  space $Q$ is Hausdorff; hence,
$Q$ is a true manifold and $V \cong \Bbb R \times Q$ (as a
manifold). \endproclaim 

\demo{Proof}
Suppose $Q$ contains a non-Hausdorff pair, $\{q,q'\}$; let
$\{\g,\g'\}$  be the corresponding foliates in $\F$.  To say
that all neighborhoods in  $Q$ of $q$ and $q'$ intersect is to
say that all tubular neighborhoods  in $V$ of $\g$ and $\g'$
have a foliate in common.  Pick points $p$ and  $p'$
respectively on $\g$ and $\g'$.  By considering fundamental 
neighborhood systems of $p$ and $p'$ and extending these (by the
$\Bbb R$-action) to tubular (i.e., $\Bbb R$-invariant)
neighborhoods, we deduce the existence of a family  of foliates
$\{\g_n\}$ and numbers $\{t_n\}$ and $\{t_n'\}$ such that 
$\{\g_n(t_n)\}$ approaches $p$ and $\{\g_n(t_n')\}$ approaches
$p'$.

As in the proof of Lemma 1.1, we form flow boxes $W$ and $W'$
around,  respectively, $p$ and $p'$; since $V$ is Hausdorff, we
can do this so  that $W$ and $W'$ do not intersect.  As in that
proof, we observe that  for $n$ large enough, $\g_n$ enters both
the past and future of $p$  within $W$, and of $p'$ within $W'$;
say $\g_n(t_n + \delta_n) \gg p \gg  \g_n(t_n - \delta_n)$ and
$\g_n(t_n' + \delta_n) \gg p' \gg \g_n(t_n' -  \delta_n)$.  Now
either, for infinitely many $n$, $t_n > t_n'$, or, for 
infinitely many $n$, $t_n < t_n'$ (or possibly both).  Assume
the former  is true; then, since the segments of $\g_n$ in $W$
and $W'$ don't overlap, we also have (for infinitely many $n$)
$t_n - \delta_n > t_n' +  \delta_n$.  Thus (for those $n$), $p
\gg \g_n(t_n - \delta_n) \gg  \g_n(t_n' + \delta_n) \gg p'$.

It follows that for all $t$ and $t'$, $\g(t)$ and $\g'(t')$ are
timelike-related.  Let $A = \{(t,t') \;|\; \g(t) \gg \g'(t')\}$
and $B = \{(t,t')  \;|\; \g(t) \ll \g'(t')\}$.  Note that $A$ and
$B$ are both open subsets of $\Bbb R^2$; since they are disjoint
and their union is the plane, it  follows that one of them is
empty, the other all of $\Bbb R^2$.  That  says precisely that
one of the two foliates is ancestral to the other. \qed\enddemo

It should be noted that an ancestral pair of leaves is not
incompatible  with a Hausdorff leaf space:  Let $V = \{(t,x) \in
\Bbb L^2 \;|\; 2x < t <  2x + 1 \}$, with $\F$ given by the 
integral curves of $\partial/\partial  t$.  Note, for instance,
that $\{(t,0) \;|\; 0<t<1\}$ is ancestral to  $\{(t,1) \;|\;
2<t<3\}$; yet the leaf space is just $\Bbb R^1$.  Whenever this
situation arises---that $\Cal F$ contains an ancestral pair,
but $Q$ is Hausdorff---any lift of $Q$ in $V \cong \Bbb R \times
Q$ is of necessity not achronal.

In spite of its being a condition that is stronger than
necessary, the absence  of ancestral pairs is a very convenient
criterion to use.  For instance,  here is a property of a
foliation $\F$ (to be used in the next section)  that implies no
ancestral pairs in $\F$ (so that Theorem 1.2 can be invoked): that
each foliate in $\F$ enter the future of each point in $V$.  If we
think of the foliate as an  observer, then this says the observer
(eventually) sees every event in  the spacetime.  Thus, let us
call a timelike curve $\g$ in a spacetime  $V$ {\it
past-omniscient\/} if $I^-(\g) = V$ (and {\it
future-omniscient\/}  if $I^+(\g) = V$); in the usage of 
\cite{HE}, these are the same, respectively, as having no future
and no past event horizon.  We will call a timelike foliation
past-omniscient  (respectively, future-omniscient) if all the
foliates are past-omniscient (respectively, future-omniscient);
we call it {\it half-omniscient\/} if it is either past- or
future-omniscient.  (Full omniscience---being both past- and
future-omniscient---will be used in Section 3.)

A simple example of a half-omniscient spacetime:  Consider the
warped product spacetime $(\alpha, \omega) \times K$ for $-\infty
\le \alpha < \omega \le \infty$ and $(K,h)$ any Riemannian
manifold, with metric $-(dt)^2 + r(t)^2 h$ for some positive
function $r$.  (For instance, Robertson-Walker spaces are of
this form, with $(K,h)$ of constant curvature.)  Let
$U = \partial/\partial t$; then $U$ is future-complete (i.e.,
its orbits extend to $\infty$ in the future, taken to be the
positive-$t$ direction) if and only if $\int_c^\omega 1/r =
\infty$ for some $c$ between $\alpha$ and $\omega$.  With $\Cal
F$ the foliation given by the orbits of
$U$, $\Cal F$ is past-omniscient if and only if $U$ is
future-complete.  (This is most easily seen in the conformally
equivalent metric $-(dt/r(t))^2 + h$, which yields a simple
static spacetime; omniscience is conformally invariant, so this
metric provides an easy venue to check on it.)

\proclaim{Proposition 1.3}
Let $V$ be a chronological spacetime and let $\F$ be a
foliation of $V$  by timelike curves.  If $\F$ is
half-omniscient, then it has no ancestral pairs. \endproclaim

\demo{Proof}
Assume $\F$ is past-omniscient.  Let $\g$ and $\g'$ be any two 
foliates.  If $\g$ is ancestral to $\g'$, then consider any
point $p'$  on $\g'$.  Being past-omniscient, $\g$ contains some
point $p \gg p'$;  however, by ancestry, $p \ll p'$, violating
chronology.  Thus, $\F$ has  no ancestral pairs. \qed\enddemo

Although, as demonstrated by the example following Theorem 1.2,
non-ancestry of foliates is not equivalent to Hausdorffness of
the leaf space, it is interesting to note that non-ancestry is
equivalent to Hausdorffness in a larger curve space, one which
contains the leaf space: the space $\Cal C(V)$ of smooth endless
causal paths (i.e., unparametrized curves) of $V$ equipped with
an appropriate topology. If $V$ is a strongly causal spacetime,
then the compact-open topology is appropriate, but for a
space-time satisfying only the chronological condition, more
subtlety is required.

The appropriate topology to use for $\Cal C(V)$ may be called
the {\it interval topology\/}:  For any collection of open sets
$\{W_1,...,W_n\}$ in $V$, let $[W_1,...,W_n] = \{c \in \Cal
C(V) \;|\;$ for some interval $J$ of $c, J \subset \bigcup
W_i$ and, for all $i, J$ intersects $W_i\}$ (by interval is meant
a connected subset of a path); then the collection of all
possible $[W_1,...,W_n]$ is a sub-basis for the interval
topology.  Note that this is, in general, finer than the
compact-open topology (which has as a sub-basis all such sets
as these with $n = 1$).  However, if $V$ is strongly causal, then
the two topologies coincide.  This is demonstrated by showing that
in a strongly causal spacetime, convergence of paths in the
compact-open sense implies convergence through connected
intervals, as can be seen by refining a chain of open sets to
a chain of simply interlocking causally convex sets: 

For any $c \in [W_1,...,W_n]$, we can cover $J$ (the relevant
interval of $c$) with a finite chain of causally convex open sets
$U_1,...,U_m$ such that each $U_i$ intersects $J$ in some
$c(s_i)$ with $s_i < s_{i+1}$, each $U_i
\cap U_{i+1} \neq \emptyset$, each $U_i\cap U_j = \emptyset$ if
$|i-j| > 1$, $\bigcup U_i \subset \bigcup W_j$, and each $W_j$
contains some $U_i \cap U_{i+1}$.  For each $i < m$, $U_i\cap
U_{i+1}$ contains a point of $c$ (since $c$ must enter
$U_{i+1}$, and only $U_i$ can contain that point of entry);
thus, $c \in \bigcap [U_i \cap U_{i+1}]$, an open set in the
compact-open topology.  Furthermore, consider any $c' \in \bigcap
[U_i \cap U_{i+1}]$: With $c'(t_i)$ the point in $U_i \cap
U_{i+1}$, on the interval $[t_i,t_{i+1}]$,  $c'$ remains within
$U_{i+1}$; thus, on $[t_1,t_{m-1}]$, $c'$ remains within
$\bigcup U_i$.  Therefore,  $\bigcap [U_i \cap U_{i+1}] \subset
[U_1,...,U_m]$, which is a subset of $[W_1,...,W_n]$.  

Thus, the following proposition is an extension of \cite{L,
Proposition 4.3}, which treats the strongly causal case in the
compact-open topology.  Theorem 1.2 tells us that non-ancestry of
a pair of foliates implies they are Hausdorff separated in
$Q$.  This proposition embeds $Q$ in $\Cal C(V)$ (so that two
elements of $Q$ which are Hausdorff separated in $\Cal C(V)$ are
{\it a fortiori}\/ Hausdorff separated in $Q$); and it also tells
us tells us that non-ancestry of a pair of foliates is equivalent
to their being Hausdorff separated in that larger space.

\proclaim{Proposition 1.4}
For any chronological space-time $V$, $\Cal C(V)$ is
Hausdorff (in the interval topology) iff it contains no ancestral
pairs, in that two curves comprise an ancestral pair if and only
if they are not Hausdorff separated.  Irrespective of ancestry,
for any foliation $\F$ of $V$ by timelike curves, the leaf space
$Q$ is topologically embedded in $\Cal C(V)$ (with the interval
topology). 
\endproclaim

\demo{Proof}
Suppose that $c$ is ancestral to $c'$. Let both curves be
parametrized by $\Bbb R$ and future-directed. For all $n$,
let $c_n$ consist of the concatenation of $c|_{(-\infty,n]}$,
a future-timelike curve from $c(n)$ to $c'(-n)$, and
$c'|_{[-n,\infty)}$, smoothed out at the corners.  Then in 
the interval topology, both $c$ and $c'$ are
limits of the sequence $\{c_n\}$; hence, $\Cal C(V)$ is not
Hausdorff.
 
Conversely, let $\{c, c'\}$ be a non-Hausdorff pair in
$\Cal C(V)$. Let $p$ and $p'$ be any points on $c$ and $c'$
respectively, and let $\{W_k\}$ and $\{W'_k\}$ be
respective (disjoint) fundamental neighborhood systems.  
Let $W_k{}^+$ denote $W_k \cap I^+(p)$, and similarly for
$-$ and primed.  For each $k$ there is a $c_k \in [W_k{}^-,
W_k, W_k{}^+] \cap [W'_k{}^-, W'_k, W'_k{}^+]$.  Each $c_k$
passes from the past to the future of $p$ before (or after)
encountering $W'_k$, and similarly for $p'$ and $W_k$.
Therefore, the same procedure as used in the proof of
Theorem 1.2 shows that $c$ and $c'$ form an ancestral pair.

Let $\pi: V \to Q$ be the natural projection, and let $j:
Q \to \Cal C(V)$ be the mapping sending $\pi (p)$ to the 
foliate $p$ lies on.  The quotient topology on $Q$ consists
of all $\pi (W)$ for $W$ a tubular open set in $V$.  If $W$
is any tubular open set in $V$, then $j(\pi (W))$ is
precisely $[W] \cap j(Q)$.  Conversely:  For any collection
of open sets $W_1,...,W_n$ in $V$, for any foliate $\g \in
[W_1,...,W_n]$, there is a tubular neighborhood $W$ of $\g$
such that, in a neighborhood of the relevant interval $J$ of
$\g$, $W$ lies inside $\bigcup W_i$.  Then $j(\pi (W))$
is a quotient-neighborhood of $\g$ lying inside
$[W_1,...,W_n]$.  \qed \enddemo

Thus a foliation $\F$ is Hausdorff as a subspace of $\Cal
C(V)$ if and only if it has no ancestral pairs.  But this really
does require the interval topology, and not the compact-open
topology, on $\Cal C(V)$ if $V$ is not strongly causal, as shown
by this example:

Let $V$ be the acausal flat cylinder: $\Bbb L^2 /\Bbb Z$, where
$\Bbb Z$ acts on $\Bbb L^2$ via  $n \cdot (t,x) = (t+n, x+n)$;
take the foliation $\Cal F$ to be given by the integral curves of
the vector field $\partial/\partial t$ (which is invariant under
the action).  Although this spacetime has closed null curves,
it is chronological.  Another way of viewing the spacetime is as
the region $\{(t,x) \;|\; 0 \le x \le 1\}$ of $\Bbb L^2$ with
$(t,0)$ identified with $(t+1,1)$; then the foliates are lines
$\{x = x_0\}$ for $0 \le x_0 < 1$.  This easily shows that there
are no ancestral pairs, and that the leaf space
$Q$ is a circle, $\Bbb S^1$.  However, $\F$ is not Hausdorff as a
subspace of $\Cal C(V)$ in the compact-open topology (as
modified for a space of unparametrized paths):  

Let $\gamma$ be the $\{x = 0\}$ foliate and let $\gamma'$ be the
$\{x = .5\}$ foliate.  In the (modified) compact-open topology, a
typical neighborhood of $\gamma$ would be the set of all causal
curves which pass through a given open set that $\gamma$ passes
through---say, for the point $p = [(0,0)]$ (where $[\,]$ denotes
equivalence class), all causal curves passing through some
neighborhood $U$ of $p$. Consider a similar
neighborhood $U'$ of a point $p' = [(.5,t_0)]$ on $\gamma'$; we
must show that there is a causal curve $c$ which intersects both
$U$ and $U'$.  Let $q'_1$ be the point on
$\gamma'$ just to the null-future of
$p$, and $q'_2$ the point on $\gamma'$ just to the null-past of
$p$, i.e., $q'_1 = [(.5,.5)]$ and $q'_2 = [(.5,-.5)]$.  If $p'$
lies to the future of $q'_1$ (i.e., $t_0>.5$), then $p \prec q'_1
\ll p'$, so $p \ll p'$, so there is a future-timelike curve $c$
from $p$ to $p'$; and $c$ clearly intersects both $U$ and $U'$. 
Similarly if $p'$ occurs on $\gamma'$ to the past of $q'_2$: $p
\succ q'_2 \gg p'$, and there is a past-timelike curve from $p$ to
$p'$.  If $q'_2 \prec p' \prec q'_1$ ($|t_0| \le .5$), then let
$c$ be the curve $[\{t = 2(1-t_0)(x-1)+1\}]$ for $.5 \le x \le
1$; this is a future-causal curve from $p'$ to $p$.  Thus, in
all cases, an element of $\Cal C(V)$ lies in both neighborhoods.  

\head
2. Causal Monotonicity
\endhead

We now consider the nature of the timelike vector field
$U$ which generates the foliation $\F$.  We will
introduce a condition on $U$ that ensures that no two
foliates are ancestrally related, and that therefore the
spacetime $V$ is topologically $\Bbb R \times Q$; this
will extend the result of Corollary 1 in \cite{H1}, which
says that if a chronological spacetime with spacetime
metric $g$ has a complete timelike vector field $U$
satisfying $\Cal L_Ug = \lambda g$ for $\lambda$
non-negative (or non-positive), then it has that same
topological form.

In the sequel, we will use $\{R_t\}$ to denote the flow of
the vector field $U$, and $t \cdot p = R_t(p)$ (as
indicated in section 1) for the point obtained by
travelling a parameter distance $t$ along the integral
curve through a point $p$ in $V$.

We want to generalize the behavior of a timelike
conformal-Killing vector field, in its effect on causal
curves; it turns out that this can be modeled fairly well
by the infinitesimal effect of the field on the metric. 
Note that a conformal-Killing field has a conformal flow,
which preserves causal character; for infinitesimal
effects, note that the Lie derivative of the metric along
such a field is proportional to the metric, so vanishes on
any null vector.  The generalizations of these properties
in either of two directions (past or future) will be
called ``causal monotonicity"; for generalization in just
one direction we will use the terms ``causally
decreasing'' or ``causally increasing'':

\definition{Definition} A diffeomorphism $f: V \to V$ is
{\it causally decreasing\/} if it maps any causal curve to
a causal curve and any timelike curve to a timelike one
(preserving past/future), and {\it strictly causally
decreasing} if it maps any causal curve to a timelike
curve (preserving past/future).
\enddefinition

It should be clear that the properties above are
equivalent to the differential $f_*$ mapping causal
vectors to causal, and so on.  Furthermore, this is
equivalent to $f_*$ just carrying null vectors to,
respectively, causal or timelike vectors:  A
future-timelike vector $X$ can be characterized as a
vector which is expressible as a linear combination
$aN+bL$ where $N$ and $L$ are future-null vectors and $a$
and $b$ are positive; thus, $f_*N$ and $f_*L$
future-causal implies $f_*X$ future-timelike.

\definition{Definition} Let $g$ be the spacetime metric,
with the convention $g(X,X) \leq 0$ for causal vectors
$X$; a vector field $U$ on $V$ is {\it causally
decreasing\/} if for any null vector $N$, $(\Cal
L_Ug)(N,N) \leq 0$, and {\it strictly causally decreasing}
if for any null vector $N$, $(\Cal L_Ug)(N,N) < 0$.
\enddefinition

It was shown in \cite{H2} that behavior of a timelike
vector field in the non-strict sense above is equivalent
to the same in that of its flow; in the strict sense, the
behavior of the flow yields the behavior of the vector
field (in that article, ``monotonic'' was used where here
``decreasing'' is used).  For completeness, we restate this
result and sketch the proof here:

\proclaim{Theorem 2.1} A timelike vector field $U$ is
causally decreasing iff its flow $\{R_t\}$ in the forward
direction (i.e., $t>0$) is causally decreasing; if the
vector field is strictly causally decreasing, then so is
its forward flow.
\endproclaim

\demo{Proof} ItÕs easy to see that if the flow of $U$ is
causally decreasing, then so is $U$:  For any null vector
$N$ at $p$, just consider the extension of $N$ to a
flow-invariant field along the integral curve $\g_p$;
then $(\Cal L_Ug)_p(N,N) = (d/dt)g(N_t,N_t)\;|\;_{t=0}$ (where
$N_t = N_{\g_p(t)}$).  If the flow is causally
decreasing, then $N$ remains causal, so that derivative is
nonpositive, giving the result.

For the converse, let $N$ be an arbitrary null vector at a
point $p$, and again extend this to a flow-invariant field
along $\g_p$.  First, let us assume the stronger
hypothesis, that $U$ is strictly causally decreasing.  As
before, $(\Cal L_Ug)_t(N,N) = (d/ds)(g(N_s,N_s))\;|\;_{s=t}$;
thus, $U$ strictly causally decreasing at $0$ (where $N$
is null) implies $N_t$ is timelike for small $t>0$ and
spacelike for small $t<0$.  In fact, $N_t$ must be
timelike for all $t>0$, since otherwise there is some
first point $t_0>0$ at which it is null, and the same
argument then applies to $N_{t_0}$---but $N_{t}$ cannot be
spacelike for $t$ just less than $t_0$.  (This establishes
the last clause in the theorem.)

Next, assume only that $U$ is causally decreasing.  Then
we can approximate $U$ by a sequence of strictly causally
decreasing vector fields, $\{U_n = (1 + (1/n)\lambda)U\}$
for an appropriate scalar field $\lambda$ (we just need it
to have a timelike gradient).  Then the flow of $U$ is
approximated by the flow of $U_n$, and the latter being
causally decreasing implies the same for the former. \qed
\enddemo

We define {\it causally increasing\/} for a timelike
vector field $U$ to mean $(\Cal L_Ug)(N,N) \allowmathbreak \geq
0$, with $>$ for the strict version (equivalently: $-U$ is
(strictly) causally decreasing); and for a transformation
of $V$ to mean its inverse is causally decreasing (same
for strict version).  We call a vector field or its flow
({\it strictly\/}) {\it causally monotonic\/} if it is
either (strictly) causally decreasing or (strictly)
causally increasing.  We then obviously have

\proclaim{Corollary 2.2} A timelike vector field $U$ is
causally monotonic iff its forward flow $\{R_t\}$ is
causally monotonic; if the vector field is strictly
causally monotonic, then so is its forward flow. \qed
\endproclaim

We note that strict causal monotonicity of the flow does
not, in general, imply the same for the vector field: 
Consider the manifold $M = \{(t,x) \in \Bbb R^2 \;|\; t>-1\}$
with metric $g = (dx)^2 - (t^3+1)(dt)^2$, and with the
vector field $U = \partial/\partial t$.  Then the 
forward flow of $U$ carries causal vectors to timelike
vectors (one need check only the null vectors), so
this is strictly causally decreasing.  However, $\Cal
L_Ug = -3t^2(dt)^2$, which vanishes at $t=0$, so $U$
is only causally decreasing, and not strictly so.

It is the causal monotonicity of the flow of a vector
field which yields the desired global property of the
corresponding foliation (by the integral curves of the
vector field).  In light of Corollary 2.2, this local
property of the flow is easily detected by the causal
monotonicity of the vector field, an infinitesimal
property.  As was shown in Theorem 7 of \cite{H2}, 
the explicit connection is this:

\proclaim{Theorem 2.3} If $U$ is a causally monotonic 
and complete timelike vector field in a chronological
spacetime $V$, then the foliation $\F$ of integral
curves of $U$ is half-omniscient. (In particular, if
$U$ is future-directed and causally decreasing, then
$\F$ is past-omniscient.) 
\endproclaim

\demo{Proof} (slightly simplifying the proof in \cite{H2}) Let $p$
be any point in $V$ and $\g: \Bbb R \to V$ any integral curve of
$U$; with $U$ future-directed  and causally decreasing, we need
to show $\g$ enters the future of $p$.  Let $\sigma: [0,1] \to V$
be a curve from $p$ to $\g(0)$, and define $\alpha: \Bbb R
\times [0,1] \to V$ by $\alpha(t,s) = t \cdot
\sigma(s)$; let $T = \alpha_*(\partial/\partial t)$
(this is actually $U$) and $S =
\alpha_*(\partial/\partial s)$ (so both $T$ and
$S$ are flow-invariant).  For some constant $m>0$,
$(mT+S)_{(0,s)}$ is future-timelike for all $s \in
[0,1]$.  It then follows from Theorem 2.1 that for all
$t>0$, $(mT+S)_{(t,s)} = {R_t}_*(mT+S)_{(0,s)}$ is
timelike.  Thus, the integral curve $\tau$ of $mT+S$,
$\tau(t) = \alpha(mt,t)$, is future-timelike for $t 
\geq 0$.  In particular, $\tau(1) = \g(m)$ is to the
future of $\tau(0) = p$.  \qed \enddemo

Finally, we obtain a result which substantially
generalizes Corollary 1 of \cite{H1} (announced as Theorem 4 of
\cite{HL}):

\proclaim{Corollary 2.4} Let $V$ be a chronological spacetime with
a causally monotonic and complete timelike vector field $U$, and
let $Q$ be the space of integral curves of $U$;
then $Q$ is a manifold and $V \cong \Bbb R \times Q$.
\endproclaim

\demo{Proof} Immediate from Theorem 1.2, Proposition 1.3,
and Theorem 2.3. \qed \enddemo

\head
3. Omniscience, the Shape of Space, and Conformastationary
Spacetimes
\endhead

The main burden of this paper is to outline some common
situations in which there is a well-defined ``shape of space". 
The key notion is to think of an edgeless spacelike hypersurface
as exemplifying a possible shape of space; if all such
hypersurfaces must be diffeomorphic to one another, then the use
of the definite article is justified:  There is only one
topology possible for something representing all of space.  But
what should ``edgeless" mean in this regard?  A number of
definitions are possible, but the easiest one to work with is
that of a properly embedded hypersurface.

Recall that if $i: M \to V$ is injective, then $i$ is
proper if and only if for any sequence $\{x_n\}$ in $M$, if
$\{i(x_n)\}$ is a convergent sequence in $V$, then $\{x_n\}$
converges in $M$.  $M$ is a spacelike hypersurface if, for
$g$ being the spacetime metric on $V$, the pulled-back
metric $i^*g$ is a Riemannian metric on $M$; and $M$ is acausal
if for no two points $x$ and $y$ of $M$, is $i(x)$ in the causal
past of $i(y)$.

\definition{Definition} By a {\it potential shape of space\/} for a
spacetime $V$ we will mean a properly embedded hypersurface $S
\subset V$ such that any acausal, properly embedded, spacelike
hypersurface (not necessarily connected) in $V$ must be
diffeomorphic to $S$.  We will call a potential shape of space for
$V$ an {\it actual shape of space\/} if we know that there
actually exists an acausal, properly embedded, and spacelike
hypersurface in $V$.
\enddefinition 

(We will see in section 4 that this definition is, in a sense,
stronger than it appears: If ``properly embedded" for $M$ is
weakened to ``edgelessly immersed" using either of a couple of
suitable notions of ``edgeless", then the definition is, in
actuality, unchanged, as any such edgeless immersion must actually
be a proper embedding.)

Note that we are not insisting that a merely potential shape of
space for $V$ be itself embeddable as an acausal or even spacelike
hypersurface---just as an edgeless hypersurface.  Clearly,
any two actual shapes of space for a given spacetime must be
diffeomorphic to one another, though that is not evident for a
potential shape of space.  But it seems that the weaker notion is
one that occurs naturally.

Here is an example of a chronological (but not strongly causal)
spacetime with a potential, but not actual, shape of space:

Put slits in $\Bbb R^2$ by deleting the following line segments: 
for every even number $2n$, each closed vertical interval of
length 1 along $x = 2n$ with lower end at $y = $ an odd number;
and for every odd number $2n+1$, each closed vertical interval of
length 1 along $x = 2n+1$ with lower end at $y =$ an even number. 
Call this slit plane, with the Euclidean metric, $N$. 
Consider the $\Bbb Z$-action on the spacetime $P =\Bbb L^1 \times
N$ defined by $(t.x,y)\cdot n = (t+2n,x+2n,y)$; since no
curve in $N$ between $(x,y)$ and $(x+2,y)$ has length as small as
2, this action does not move any point to one in its past or
future.  Thus $V = P/\Bbb Z$ is (just barely) chronological; it
is not strongly causal.  The integral curves of
$\partial/\partial t$ in $P$ form a timelike foliation, preserved
by the $\Bbb Z$ action; $V$ inherits this foliation.  The leaf
space in $P$ is $N$; in $V$ it is $Q = N/\Bbb Z$.  As $V$ is
static-complete, Corollary 3.2 (below) tells us that $Q$ is a
potential shape of space for $V$.  But although $Q$
can be properly embedded in $V$ via $[x,y] \mapsto [x,y,x]$ (square
brackets denoting equivalence class under the group action), there
is no acausal proper embedding of $Q$ into $V$: $Q$ is not an
actual shape of space for $V$, and so its status as a potential
shape of space is vacuous.

\remark{Conjecture} Section 4 will illustrate some of the stronger
results obtainable when one insists on strong causality, not
merely chronology.  Perhaps that is the key element for a strong
shape of space?  This seems to be a reasonable conjecture:  A
potential shape of space for a strongly causal spacetime must be an
actual shape of space.
\endremark
\medpagebreak

The most salutary virtue of an omniscient foliation is
that it provides a potential shape of space (as announced in
\cite{HL} as Theorem 2, where it was called simply ``shape of
space"):

\proclaim{Theorem 3.1}
Let $V$ be a chronological spacetime with an omniscient foliation
$\Cal F$ of timelike curves.  Then the leaf space $Q$ is a
potential shape of space for $V$.
\endproclaim

\demo{Proof} By Proposition 1.3 and Theorem 1.2, $Q$ is
Hausdorff.  As argued above (before Theorem 1.2), this implies
that the line-bundle $\pi: V \to Q$ has a cross-section $\sigma: Q
\to V$ (in the smooth category), and $\sigma$ must be a proper
embedding:  Since $\pi \circ \sigma = 1_Q$, $\sigma$ is an
injective immersion; if $\{\sigma(q_n)\}$ converges to some $p
\in V$, then $\{\pi(\sigma(q_n))\} = \{q_n\}$ converges to
$\pi(p)$.  (But there is no guarantee that $\sigma$ is spacelike
or, if it is, that $\sigma(Q)$ is acausal.)

Consider any acausal, spacelike, proper embedding $i: M \to V$,
with dim$(M)$ = dim$(V) - 1$; we will show that $\pi \circ i: M
\to Q$ is a diffeomorphism.

Since $i(M)$ is acausal, $\pi \circ i$ must be injective (else two
points of $M$ will be mapped by $i$ to the same foliate).  Let
$g$ be the metric on $V$; since $i^*g$ is Riemannian, $\pi \circ
i$ is an immersion ($\pi_*$ kills only timelike vectors).  Then,
by invariance of domain, $(\pi \circ i)(M)$ is an open subset of
$Q$, and $\pi \circ i$ is a diffeomorphism onto this image; all
we need show is that $\pi \circ i$ has a closed image.

Let $\{x_n\}$ be a sequence in $M$ with $\{\pi(i(x_n))\}$
converging to some point $q \in Q$.  Let $\gamma_n$ be the
foliate corresponding to $\pi(i(x_n))$ and $\gamma$ the foliate
corresponding to $q$, parametrized so that they begin
in $\sigma(Q)$: $\gamma_n(0) =
\sigma(\pi(i(x_n)))$ and $\gamma(0) = \sigma(q)$.  We can define
a continuous function $\tau : V \to \Bbb R$ by measuring how far
each point of $V$ is from $\sigma(Q)$, in terms of the real
action:  For any $p \in V$, $\tau(p) \cdot \sigma(\pi(p)) = p$. 
If we can show that the sequence $\{\tau(i(x_n))\}$ has a
convergent subsequence, then we will be done:  If
$\{\tau(i(x_{n_k}))\}$ converges to $t$, then $\{i(x_{n_k})\} =
\{\tau(i(x_{n_k})) \cdot \sigma(\pi(i(x_{n_k})))\}$ converges to
$t \cdot \sigma(q)$; since $i$ is proper, this means $\{x_{n_k}\}$
converges to some $x \in M$, with $i(x) = t \cdot \sigma(q)$. 
Then $\pi(i(x)) = \pi(t \cdot \sigma(q)) = q$, and $q \in
\pi(i(M))$.

To show that $\{\tau(i(x_n))\}$ has a convergent subsequence, we
will look at the points $\{i(x_n)\}$ for $n$ sufficiently large: 
Consider that $\gamma$ must enter the future of $i(x_1)$ at some
point $p^+$ in $V$.  The boundary of $I^+(i(x_1))$ is a
three-dimensional topological manifold $B^+$ through $p^+$,
transverse to foliates of $\Cal F$; accordingly, we can find a
relatively compact neighborhood $U$ of $q$ in $Q$ such that the
foliates $\gamma'$ corresponding to points $q' \in U$ each
intersect $B^+$ within a relatively compact neighborhood of
$p^+$.  We can do the same thing with $p^-$ being the
point where $\gamma$ enters $I^-(i(x_1))$ and $B^-$ the boundary
of that past: the foliates sufficiently close to $\gamma$ (we
shrink $U$ if necessary) intersect $B^-$ at points within a
relatively compact neighborhood of $p^-$.  Then we have that
portion of $\hat U = \pi^{-1}(U)$ which is in neither
$I^+(i(x_1))$ nor $I^-(i(x_1))$ as a relatively compact set $W$
between $B^+$ and $B^-$.  For $n$ sufficiently large,
$\pi(i(x_n)) \in U$, so $i(x_n) \in
\hat U$.  Furthermore, since $i(M)$ is acausal, we cannot have
$i(x_n) \gg i(x_1)$ or $i(x_n) \ll i(x_1)$:  For these
sufficiently high $n$, $i(x_n)$ must lie in $W$.  Since $W$ is
relatively compact, the numbers $\{\tau(i(x_n))\}$ must have a
convergent subsequence. \qed 
\enddemo

(This proof actually goes through with a slightly strengthened
definition of potential shape of space: that any achronal (instead
of merely acausal), spacelike, properly embedded hypersurface be
diffeomorphic to the shape of space.  Acausal is used in the
definition in order to gain desired strength for the definition
of actual shape of space, as used in the next section.)

As a class of examples, consider conformastationary spacetimes: 
This means there is a timelike vector field $U$ so that, with $g$
the metric, $\Cal L_U(g) = \lambda g$ for some scalar function
$\lambda$ (where $\Cal L$ denotes Lie derivative).  We will call
the spacetime conformastationary-complete if the
conformal-Killing field $U$ is a complete vector field.  

\proclaim{Corollary 3.2} Any chronological
conformastationary-complete spacetime has a potential shape of
space.
\endproclaim

\demo{Proof} Let $U$ be the conformal-Killing field and $g$
the metric; then $\Cal L_U(g)$ vanishes on any null vector, so
$U$ is both causally decreasing and causally increasing. 
Therefore, by Theorem 2.3, the foliation
$\Cal F$ generated by $U$ is omniscient.  By Theorem 3.1, the
leaf space of $\Cal F$ is a potential shape of space for the
spacetime.
\qed
\enddemo

A somewhat less general result was given in \cite{GH}, Theorem 2:
In a chronological stationary spacetime which is timelike or
null geodesically complete, any achronally embedded spacelike
hypersurface which is closed as a subspace of the spacetime must
be diffeomorphic to the space of stationary observers (i.e., the
leaf space of the foliation generated by the Killing field).

For a class of conformastationary-complete spacetimes (more
general than the stationary-complete spacetimes of \cite{GH}),
consider the warped product example from section 1: $(\alpha,
\omega) \times K$ with spacetime metric $g = -(dt)^2 + r(t)^2h$
for $h$ a Riemannian metric on $K$.  The property of being
conformastationary is conformally invariant, so we can instead
consider the metric $\bar g = -(dt/r(t))^2 + h = -(d\tau)^2 + h$,
where $d\tau/dt = 1/r(t)$.  Then $U = \partial/\partial\tau$ is
clearly a Killing field for
$\bar g$, hence, a conformal-Killing field for $g$.  $U$ is
complete if and only if $\int^\omega_c 1/r = \infty$ and
$\int_\alpha^c 1/r = \infty$ for some
$c$ between $\alpha$ and $\omega$.

\head
4. Two-Sheet Omniscience and Actual Shapes of Space
\endhead
    
This section will consider situations in which we hope to derive
global properties of an immersed spacelike hypersurface---such as
its being an actual shape of space for the ambient
spacetime---from as little information as possible, such as the
action of the immersion on the fundamental group.  The key ideas
are that of an immersed hypersurface being ``edgeless" (in a
more general sense than being properly embedded) and of the
spacetime having a ``timelike-contractible disk" bounding any
null-homotopic curve.  These ideas were introduced in
\cite{H2}, but will be repeated here.

We will use this general notion of edgelessness as a weaker
hypothesis for potential shapes of space.  So long as the
hypersurface is assumed to have an acausal image, it will follow
(using a result in \cite{H2}) that an analysis on the level of
the fundamental group is sufficient to determine whether
or not the immersion provides an actual shape of space.  

The more difficult trick will be to draw such a conclusion {\it
without\/} making a causality assumption on the image.  To do
that will require a slightly strengthened version of
omniscience and that the ambient spacetime be strongly
causal, not merely chronological.  Strong results from \cite{H2}
(descending from a series of earlier results in \cite{H3, H4}),
making use of timelike contractible disks, and slightly
modified here, will be used to achieve the desired goal. 

Let $M$ be a manifold of dimension 1 less than that of the
spacetime $V$, and let $i: M \to V$ be an immersion.  This is an
approximation to a spacelike hypersurface so long as $i$ induces a
Riemannian metric on $M$:  Let $g$ be the spacetime metric on $V$;
then we call $i$ a {\it spacelike immersion\/} so long as the
pulled back metric $i^*g$ is Riemannian.  

Let $i : M \to V$ be a spacelike immersion.  One way of being
convinced that $M$ is immersed in an edgeless sort of fashion is
to note that $i^*g$ is complete.  But since we are interested in
conformally invariant notions here, we will generalize this to
{\it conformally completable\/}:   That means that for some
positive function $\Omega : V \to \Bbb R^+$, $i^*(\Omega g)$ is
complete.  Another notion of edgelessness would be to have $i :
M \to V$ be proper; but that is stronger than is actually
required, as much can be proved merely from looking at curves
(this is one of the major themes in \cite{H2}).  Accordingly, it
is useful to consider the property of being {\it
curve-proper\/}:  This means that for any curve $c : [0,1) \to
M$, if its image in $V$,
$i\circ c$, has an endpoint at 1---i.e.,
$\lim_{t\to1}i(c(t))$ exists in $V$---then $c$ has (in
$M$) an endpoint at 1 also.  These notions are both strictly
weaker than proper (let $i$ map the line into an asymptotically
``horizontal" spacelike helix in the Minkowski cylinder, $i :
\Bbb R \to \Bbb L^1 \times \Bbb S^1$ with $i : x \mapsto
(\tanh(x/2),[x])$, where $[x]$ is the projection of $x$ into the
circle; then $i$ is a spacelike immersion of codimension 1 which
is curve-proper and conformally completable, but not proper). 
Either of these ideas is an acceptable notion for edgelessness:

\definition{Definition} A spacelike immersion is called {\it
edgeless\/} if it is curve-proper or conformally completable.
\enddefinition

This is the notion of edgelessness referred to in the previous
section, just after the definition of shape of space.  Theorem 2
in \cite{H2} says that for an edgeless spacelike
immersion $i : M  \to V$ of codimension 1, if the image $i(M)$ is
achronal, then $i(M)$ is actually a properly embedded
hypersurface, and $i: M \to i(M)$ is a covering projection; an
example is $i: \Bbb R \to \Bbb L^1 \times \Bbb S^1$ with $i: x
\mapsto (0,[x])$.  Thus, if $S$ is an actual shape of space for
$V$, then not only is any properly embedded, acausal, spacelike
hypersurface diffeomorphic to $S$, but also any merely immersed,
acausal, spacelike hypersurface must be as well, so long as it is
edgeless---though we must take care to speak of the image of the
immersion being diffeomorphic to $S$, as the domain may be a cover
of $S$.  (This is not really a stronger statement of the meaning
of shape of space, as any such immersed hypersurface---or, at
any rate, its image---must actually be embedded.)

\proclaim{Proposition 4.1} Let $V^n$ be a chronological spacetime
possessing an omniscient foliation $\Cal F$ by timelike curves. 
Let $i : M^{n-1} \to V$ be an edgeless spacelike immersion with
acausal image $i(M)$.  Then $i_* : \pi_1(M) \to \pi_1(V)$ is
injective; and if $i_*(\pi_1(M)) = \pi_1(V)$, then $i: M \to V$
is an actual shape of space for $V$.
\endproclaim

\demo{Proof}  By Theorem 2 of \cite{H2}, we know that $i(M)$ is a
properly embedded spacelike hypersurface.  We know from Theorem
3.1 here that $Q = V/\Cal F$ is a potential shape of space for $V$,
so $i(M)$ is diffeomorphic to $Q$; in fact, $\pi : i(M) \cong Q$
(where $\pi: V \to Q$ is projection).  Thus, $\pi_1(i(M)) \cong
\pi_1(Q) \cong \pi_1(V)$ (the last because $\pi$ is a projection
with contractible fibre).  Since $i : M \to i(M)$ is a covering
map, $i_* : \pi_1(M) \to \pi_1(i(M))$ is injective, and it is
surjective if and only if $i: M \to i(M)$ is a homeomorphism.  

Since $\pi : i(M) \to Q$ and $\pi: V \to Q$ both induce
isomorphisms of fundamental groups, we can translate the results
above into statements about $i : M \to V$:  $i_* : \pi_1(M) \to
\pi_1(V)$ is injective, and it is also surjective if and only if
$i : M \to i(M)$ is a homeomorphism.  In the latter case, $\pi
\circ i : M \to Q$ is also a homeomorphism, so $M$ is a
potential shape of space for $V$ (since $Q$ is).  Furthermore, we
then have $M$ properly embedded in $V$ via $i$ in an acausal
manner (since $i(M)$ is properly embedded and $i$ is a
homeomorphism onto its image), so $M$ is an actual shape of
space. \qed
\enddemo

Edgelessly immersed spacelike hypersurfaces are looked at in
detail in \cite{H2} in the context of spacetimes which, broadly
speaking, are not spacelike at timelike infinity.  The key idea
(in a nutshell) is that disks be suffered to exist long
enough to shrink to a point.  More precisely:  We need that any
null-homotopic loop in the spacetime be the boundary of a
disk which is timelike-contractible:

\definition{Definition} A {\it timelike-contractible disk\/} in a
spacetime $V$ is an immersion $B: (-1,1)\times D^2 \to V$, where
$D^2$ is the closed disk in the plane, such that 
\roster
\item for any $p \in D^2$, $B(-,p) : (-1,1) \to V$ is a timelike
curve, 
\item $B$ extends continuously to $[-1,1]\times D^2$, and
\item $B(1,D^2)$ and $B(-1,D^2)$ are single points.
\endroster
A timelike-contractible disk $B$ {\it spans\/} a loop $\sigma$ if
$\sigma$ is the boundary of $B(0,D^2)$.  (Somewhat more
precisely:  This is a disk in $V$---$B(0,D^2)$---together with a
timelike contraction of the disk; the disk itself need not have
any particular causal nature.)
\enddefinition

As is shown in \cite{H2}, timelike-contractible disks are the
key to having a spacetime sufficiently well behaved at timelike
infinity: sufficiently well that edgeless spacelike immersions of
codimension 1 are actually proper embeddings, so long as they do
the right thing on the level of the fundamental group;
specifically, one must have a timelike-contractible disk
spanning each null-homotopic loop.  But how can one tell if a
spacetime has this property?  Omniscient foliations of the sort
considered here provide an easy answer---almost.  We need to
strengthen the notion of omniscience just a bit:

We need to specialize our consideration of a foliation $\Cal F$
to those foliates which intersect any particular curve in the
spacetime $V$.  This can be expressed in terms of timelike
2-surfaces in $V$:  Given a non-degenerate (but possibly
self-intersecting) curve $c: (a,b) \to V$, we consider
$P_c$, roughly the union of all the foliates which contain any
$c(s)$, parametrized by $s$.  We almost can look at this as an
immersed 2-surface by considering $i: \Bbb R \times (a,b) \to V$
defined by $i(t,s) = t \cdot c(s)$; but this fails to be an
immersion where $\dot c(s)$ is parallel to the foliate through
$c(s)$.  So instead consider the projection $\pi \circ c$; so
long as $c$ is not wholly lying along a single foliate, this
curve can be parametrized as a non-degenerate curve $\delta :
(\alpha, \beta) \to Q$.  Then $P_c$ is
$\pi^{-1}(\text{Im}(\delta))$, expressed as an immersion via $i:
\Bbb R \times (\alpha,\beta)
\to V$ with $i(t,s) = t\cdot\sigma(\delta(s))$ (where $\sigma: Q
\to V$ is a cross-section of $\pi$, as before); then $P_c$ is a
timelike 2-sheet in $V$ (exceptionally: if $c$ is lies along a
single foliate $\gamma$, then $P_c$ is just $\gamma$).  The
foliation $\Cal F$ induces a foliation $\Cal F_c$ on $P_c$ (in
the exceptional case $\Cal F_c$ is just the single foliate
$\gamma$).

\definition{Definition} A foliation $\Cal F$ of timelike curves
in a spacetime $V$ will be called {\it 2-sheet omniscient\/} if
for every non-degenerate curve $c$ in $V$, not lying along a
single foliate, the induced foliation $\Cal F_c$ is omniscient
in the timelike 2-sheet $P_c$.  (This is the same as saying
that in the spacetime $\Bbb R \times (\alpha,\beta)$ with metric
$i^*g$---$g$ the spacetime metric in $V$---the foliation
$\{\gamma_s : t \mapsto (t,s)\;|\;\alpha < s < \beta\}$ is
omniscient.)  
\enddefinition

This actually is a stronger notion of omniscience, as can be
seen by considering the following spacetime:  Start with $\Bbb
L^3$ (metric $-dt^2 + dx^2 + dy^2$), but in the region $\{|y| <
1 \}$, narrow the lightcones in the $x$-direction so as create
``particle horizons" in the $y = \text{constant}$ planes,
i.e., the null curves in those planes have vertical
asymptotes.  This can be done, for instance, with metric $-dt^2
+ f(t,y)^2dx^2 + dy^2$, where 
$$f(t,y) = \left\{
\aligned
&(1-y^2)t^2+1, \\
&1, \\
\endaligned \quad
\aligned
&|y| < 1 \\
&|y| \ge 1 \\
 \endaligned
\right.\;.$$ 
Let $\Cal F$ be the foliation of $t$-curves, $\{x = x_0, y =
y_0\}$.  Then the induced foliation on each $\{y = y_0\}$
plane, with $|y_0| < 1$, is not omniscient (for instance, in
$\{y = 0\}$, the null curves through $(0,0,0)$ have
asymptotes at $x = \pm\pi/2$); but in the entire
spacetime, $\Cal F$ is omniscient, since a timelike curve
exists from any point $(t_0,x_0,y_0)$ to any foliate $\{x =
x_1, y = y_1\}$, even in the curved region
$|y| < 1$, by first traveling in $x = x_0$ to the
flat region, travelling there to $x = x_1$, then back along $x
= x_1$ to the foliate. 

It is interesting to note that this version of omniscience is
inherited by covering spaces:  If $\widetilde V$ is the universal
covering space of the spacetime
$V$, and $V$ has a  foliation $\Cal F$ by timelike curves, then
$\Cal F$ induces a foliation $\widetilde{\Cal F}$ of timelike
curves on $\widetilde V$ (lifting via $P : \widetilde V \to V$,
which is locally a diffeomorphism); and if $\Cal F$ is 2-sheet
omniscient, then so is $\widetilde{\Cal F}$.  This can be seen by
considering the timelike 2-sheet $\widetilde\Pi$ in $\widetilde V$
generated by a curve $\tilde c : [0,1] \to \widetilde V$ and trying
to find a timelike curve from $\tilde c(0)$ to the foliate
$\widetilde \gamma_1$ ($\widetilde\gamma_s$ being the foliate through
$\tilde c(s)$).  In $\Pi = P(\widetilde \Pi)$, there is a timelike
curve $\delta$ from $c(0)$ to $\gamma_1$ ($c = P \circ \tilde
c$, $\gamma_s = P \circ \widetilde\gamma_s$); and $\delta$ can be
expressed as $\delta(s) = \lambda(s) \cdot c(s)$ for some
function $\lambda: [0,1] \to \Bbb R$.  The loop $\tau_1$ formed
by $c$, $\gamma_1|_{[0,\lambda(1)]}$, and $\delta$ is
null-homotopic, as provided by the family of loops
$\tau_s$ formed by $c|_{[0,s]}$, $\gamma_s|_{[0,\lambda(s)]}$,
and $\delta|_{[0,s]}$.  Thus, the loop $\tau_1$ lifts to a loop
$\widetilde \tau_1$ in $\widetilde V$, lying in $\widetilde\Pi$, where its
timelike portion $\delta$ goes from $\tilde c(0)$ to $\widetilde
\gamma_1$.

(Ordinary omniscience is not necessarily inherited by covering
spaces:  Consider as $V$ the same spacetime as just above,
restricted to $y > 0$, and with the line $\{x = 0, y = 1/2\}$
removed.  In the universal covering space $\widetilde V$, the
induced foliation $\widetilde{\Cal F}$ is not omniscient, since
going, in $V$, from $(0, -1, 1/4)$ around the missing line to
get to the flat region, and back to the foliate
$\{x = 1, y = 1/4\}$ is not in the same homotopy class as
staying in $\{y = 1/4\}$; thus, the lift in $\widetilde V$ reaches a
different foliate.)

\proclaim{Proposition 4.2}Let $V$ be a spacetime with a
foliation $\Cal F$ by timelike curves, which is 2-sheet
omniscient.  Then for every null-homotopic loop $\sigma$ in $V$,
there is a timelike-contractible disk spanning $\sigma$.
\endproclaim

\demo{Proof} We need work only in the differentiable category.

Let $\sigma : \Bbb S^1 \to V$ be a null-homotopic loop in $V$;
then there is a disk spanning $\sigma$, which we can take to be
an immersion $b : D^2 \to V$, where $D^2$ is the closed unit disk
in the plane, with the restriction of $b$ to the boundary of
$D^2$ being
$\sigma$.  

Let $\pi : V \to Q$ be the projection to the leaf space of $\Cal
F$. In virtue of $\Cal F$ being 2-sheet omniscient, we know the
following:  For any curve $c: [0,1] \to Q$ and for any point $p
\in \pi^{-1}(c(0))$, there is a unique future-null curve $c^+_p :
[0,1] \to V$ which is a lift of $c$ starting at $p$, i.e.,
$c^+_p(0) = p$ and $\pi \circ c^+_p = c$, and also a unique
past-null lift $c^-_p$ of $c$ starting at $p$. We know this 
because the induced foliation on the 2-sheet $P = \pi^{-1}(c)$ is
omniscient, so the foliate $\gamma_1 = \pi^{-1}(c(1))$ enters
both the past and future of $p$ in terms of the Lorentz manifold
$P$ (more precisely: in terms of the Lorentz manifold $\Bbb R
\times [0,1]$, immersed with image $P$ in $V$); at some point,
it crosses the null curve in $P$ from $p$.  Furthermore, if we
allow the curve $c$ and the point $p$ to vary, then $c^+_p(1)$
and $c^-_p(1)$ vary continuously with $c$ and $p$ in the
$\text{C}^1$ topology, since $c^+_p$ and $c^-_p$ are given as
solutions to differential equations with these as input data.

Let us parametrize the disk $D^2$ by radial segments $\{r_\theta
: [0,1] \to D^2 \;|\; \theta \in \Bbb S^1\}$, with $r_\theta(0)$
on the boundary (at position $\theta$) and $r_\theta(1)$ at the
center.  For each $\theta \in \Bbb S^1$, let $c_\theta = \pi
\circ b \circ r_\theta$; then the future-null lifts of the
images of $r_\theta$, starting at the corresponding points of
$\sigma$, vary continuously, i.e., the points
$\{(c_\theta)^+_{\sigma(\theta)}(1)\}$ vary continuously in
$\theta$.  Let $\gamma$ be the foliate through $b(0)$; then
there is a point $p^+$ on $\gamma$ that is to the future of each
$(c_\theta)^+_{\sigma(\theta)}(1)$.  Similarly, there is a point
$p^-$ on $\gamma$ to the past of each
$(c_\theta)^-_{\sigma(\theta)}(1)$.

Consider each $P_\theta = \pi^{-1}(c_\theta)$; more precisely,
$P_\theta$ is $\Bbb R \times [0,1]$ with metric pulled back from
the immersion $i_\theta: \Bbb R \times [0,1] \to V$ defined via
$c_\theta$, i.e., $i_\theta(t,s) = t \cdot c_\theta(s)$.  The
points $p^+$ and $p^-$ in $V$ are represented by the same points
$(t^+,1)$ and $(t^-,1)$ in each $P_\theta$, while $(0,0)$ in
$P_\theta$ represents $\sigma(\theta)$ in $V$. There is a
future-timelike curve $\delta^0_\theta$ in $P_\theta$ from
$(t^-,1)$ through $(0,0)$ to $(t^+,1)$; we can parametrize this
as $\delta^0_\theta : [-1,1] \to P_\theta$ with
$\delta^0_\theta(-1) = (t^-,1)$, $\delta^0_\theta(0) = (0,0)$,
and $\delta^0_\theta(1) = (t^+,1)$.  Then we can fill in the
portion of $P_\theta$ between $\delta^0_\theta$ and $\Bbb R
\times \{1\}$ with a timelike foliation $\{\delta^s_\theta :
[-1,1] \to P_\theta \;|\; 0 \le s \le 1\}$ with
$\delta^s_\theta(-1) = (t^-,1)$,
$\delta^s_\theta(0) = (0,s)$, and $\delta^s_\theta(1) =
(t^+,1)$, where $\delta^1_\theta$ runs along $\Bbb R \times
\{1\}$.

(In case $c_\theta$ is degenerate---i.e., $b\circ
r_\theta$ lies along a single foliate $\gamma_\theta$---then
$P_\theta$ is just $\Bbb R$, mapped into $V$ as
$\gamma_\theta$.  In that case, 
$(c_\theta)^+_{\sigma(\theta)}$ and
$(c_\theta)^-_{\sigma(\theta)}$ are each the degenerate
curve constant at $\sigma(\theta)$. The curves $\delta^s_\theta$
are each a map taking $-1$ to $t^-$, 0 to 0, and 1 to $t^+$.) 

We can form these various $\delta^s_\theta$ in a manner that
is differentiable in $\theta$.  Putting them all together
then yields the map $B: [-1,1]\times[0,1]\times\Bbb S^1
\to V$ given by $B(t,s,\theta) = i_\theta(\delta^s_\theta(t))$,
which is an immersion except at the various boundaries.  (In
case some $c_\theta$ is degenerate---which can happen only
for isolated values of $\theta$---we can vary the map $B$ in a
neighborhood $[-1,1]\times[0,1]\times\{\theta\}$ so as to
preserve its status as an immersion, while keeping
$B(-,s,\theta)$ a timelike curve from $p^-$ to $p^+$.) Clearly,
$B$ is a timelike contraction of a disk spanning $\sigma$.
\qed 
\enddemo

All we need do now to invoke the theorems of \cite{H2} is 
upgrade the chronology assumption on $V$ to strong causality.

\proclaim{Theorem 4.3}Let $V^n$ be a strongly causal spacetime
possessing a 2-sheet omniscient foliation $\Cal F$ of timelike
curves.  Let $i: M^{n-1} \to V$ be an edgeless spacelike
immersion.  Then $i_* : \pi_1(M) \to \pi_1(V)$ is
injective; and if $i_*(\pi_1(M)) = \pi_1(V)$, then $i: M \to V$
is an actual shape of space for $V$.
\endproclaim

\demo{Proof} By Proposition 4.2, every null-homotopic loop in
$V$ is spanned by a timelike-contractible disk.  Corollary 5 of
\cite{H2} says precisely that for a strongly causal spacetime
$V$ with that property, for any codimension-1 edgeless spacelike
immersion $i: M \to V$, $i_* : \pi_1(M) \to \pi_1(V)$ is
necessarily injective; and that if $i_*$ is surjective, then $i$
must be an achronal proper embedding.

In fact, $M$ is diffeomorphic to the leaf space $Q$ of the
foliation $\Cal F$:  Theorem 3.1 tells us $Q$ is a potential shape
of space for $V$ and, in fact, that $\pi \circ i : M
\to Q$ is a diffeomorphism (the parenthetical comment directly
following Theorem 3.1 indicates that we need have $M$ only
achronally embedded, not acausally).

Suppose $N$ is any acausal and properly embedded hypersurface
in $V$; then another application of Theorem 3.1 shows $N$ is
diffeomorphic (via $\pi$) to $Q$, hence, to $M$.  This shows
$M$ is a potential shape of space for $V$ (but since we have not
yet shown $M$ to be acausally embedded, we cannot yet say it is an
actual shape of space).

All that remains is to show that $M$ is, indeed, acausally
embedded in $V$, not just achronally.  This requires a slight
strengthening of the results in \cite{H2} and its predecessors
\cite{H3, H4}.  This is essentially technical in nature,
involving one small piece of substantive work and a deal of
minor bookkeeping.  

Specifically, what is needed is to strengthen Corollary 5
of \cite{H2}, which says that if $V^n$ is strongly causal and
has every null-homotopic loop spanned by a timelike-contractible
disk, then any edgeless spacelike immersion $i: M^{n-1} \to V$
must be injective on the level of the fundamental group, and if
$i_* : \pi_1(M) \to \pi_1(V)$ is onto, then $i$ must be
an achronal proper embedding.  This must be strengthened to
say that that $i$ must be acausal, not merely achronal. 
Backtracking through \cite{H4} and \cite{H3}, one sees that
the crucial step comes in Theorem 3 of \cite{H3}.  

An argument is given there that says that in a Lorentz manifold
homeomorphic to the plane, a timelike and a
spacelike curve cannot intersect twice (since from any given
point $p$, the null cone from $p$ separates the timelike curves
issuing from $p$ from the spacelike curves issuing from $p$). 
But this argument applies equally well to a causal and a
spacelike curve.  That observation then propagates through
Proposition 4 and Theorems 5 and 6 of \cite{H3},
Proposition 1, Theorem 2, and Corollary 3 of \cite{H4}, and
Theorem 3, Corollary 4, and Corollary 5 of \cite{H2} (with
appropriate changes being made in a handful of definitions
along the way).

The strengthened Corollary 5 of \cite{H2} then
does the trick here:  From Proposition 4.2, we know every
null-homotopic loop in $V$ is spanned by a timelike-contractible
disk.  Thus, $i$ embeds $M$ into $V$ in an acausal manner. \qed 
\enddemo

As an application of the ideas from this and the previous
section, consider strongly causal, conformastationary-complete
spacetimes:  We know from Corollary 3.2 that the
conformastationary-observer space is a potential shape of space for
such a spacetime; but with strong causality, more information is
available to us (this was announced as Theorem 3 in \cite{HL}):

\proclaim{Theorem 4.4} Let $V^n$ be a strongly causal,
conformastationary-complete spacetime, and let $Q$ be the
conformastationary-observer space, i.e., the leaf space of the
foliation defined by the conformal-Killing vector field, with
$\pi : V \to Q$ the projection.  Let $i: M \to V$ be an
edgelessly immersed spacelike hypersurface.  Then $\pi \circ i
: M \to Q$ is a covering projection with fibre
$\pi_1(V)/i_*(\pi_1(M))$; in particular, if $i_*$ is onto,
then $M$ is homeomorphic to $Q$, and $M$ is acausally embedded.
\endproclaim   

\demo{Proof} As indicated in the proof of Corollary 3.2, we
know that the foliation $\Cal F$ defined by the
conformal-Killing field is omniscient.  But to apply Theorem
4.3, we need to show it is 2-sheet omniscient.  So let $c$ be
any non-degenerate curve in $V$ (not consisting of only a single
conformastationary-observer orbit), and let $P_c$ be the
immersed timelike 2-surface defined by the orbits passing
through $c$ (essentially, $P_c = \pi^{-1}(\pi(c))$), with $\Cal
F_c$ the induced foliation from $\Cal F$ and $g_c$ the induced
metric (in detail: $\pi \circ c$ is given by a
non-degenerate curve $\delta : (\alpha, \beta) \to Q$, $P_c =
\Bbb R \times (\alpha, \beta)$, we have an immersion $j_c : P_c
\to V$ via $j(t,s) = t \cdot \sigma(\delta(s))$ for some chosen
cross-section $\sigma : Q \to V$ of $\pi$, $\Cal F_c$ is the
family $\{\Bbb R \times \{s\} \;|\; \alpha < s < \beta\}$, and
$g_c = (j_c)^*(g)$, where $g$ is the spacetime metric on $V$). 
Then $(P_c, g_c)$ is a conformastationary-complete spacetime in
its own right; thus, as in Corollary 3.2, $\Cal F_c$ is
omniscient in $P_c$.

Thus, Theorem 4.3 applies to $V$:  We know $i_* : \pi_1(M) \to
\pi_1(V)$ is injective, and that if $i_*$ is onto, then $i$ is
an acausal embedding and, by Theorem 3.1, $\pi \circ
i : M \to Q$ is a homeomorphism.  

But what happens if $i_*$ is not onto?

Let $\widetilde V$ be the universal covering space of $V$, with
$p_V: \widetilde V \to V$ the projection; then
$\widetilde V$ is also conformastationary-complete.  With $\widetilde
M$ the universal covering space of $M$ (with projection $p_M :
\widetilde M \to V$), we have an induced immersion $\tilde i: \widetilde
M \to \widetilde V$ ($p_V \circ \tilde i = i \circ p_M$).  Clearly,
$\tilde i$ inherits being edgeless and spacelike; thus, by
Theorem 4.3, it is an acausal embedding.

We have the following maps:  

The map $i : M \to V$ is covered by the map $\tilde i: \widetilde
M \to \widetilde V$, in the sense that $p_M : \widetilde M \to M$ and
$p_V : \widetilde V \to V$ are both principal covering maps
(i.e., the fibre is a group and projection is by
identification under the group action) with structure groups,
respectively, $\pi_1(M)$ and $\pi_1(V)$, and that the group
actions are preserved through the action of $i_* : \pi_1(M) \to
\pi_1(V)$, i.e., for $g \in \pi_1(M)$ and
$\widetilde x \in \widetilde M$, $i(p_M(g \cdot \widetilde x)) = p_V((i_*g)
\cdot \tilde i(\widetilde x))$.    

The projection to observer-space $\pi : V \to Q$ has its
mirror, projection to observer-space $\widetilde\pi : \widetilde V \to
\widetilde Q$.  The principal covering map $p_V : \widetilde V \to V$ is
easily seen to induce a map $p_Q : \widetilde Q \to Q$, which is
also a principal covering map; since $\widetilde Q$ is simply
connected, $p_Q$ must actually be the universal covering map for
$Q$ with structure group $\pi_1(Q)$.  Thus, $\widetilde \pi$
covers $\pi$ in the sense that for $g \in \pi_1(V)$ and $\widetilde
x \in \widetilde V$, $\pi(p_V(g \cdot \widetilde x)) = p_Q(\pi_*(g)
\cdot \widetilde\pi(\widetilde x))$. (Both $\pi$ and
$\widetilde\pi$ are principal fibre bundles with structure group
$\Bbb R$, and $p_Q(\widetilde\pi(t \cdot \widetilde x)) = \pi(t \cdot
p_V(\widetilde x))$.)

From Theorem 4.3 applied to $\tilde i$, we know $\widetilde \pi
\circ \tilde i : \widetilde M \to \widetilde Q$ is a homeomorphism. 
Thus, the map $p = p_Q \circ (\widetilde \pi \circ \tilde i) :
\widetilde M \to Q$ is a universal covering map for $Q$; since
$\tilde i$ covers $i$ and $\widetilde \pi$ covers $\pi$, we know
$\widetilde \pi \circ \tilde i$ covers $\pi
\circ i$, i.e., $p$ factors as $(\pi \circ i) \circ p_M$.  The
action of the map $p_M$ is to factor out the action of
$\pi_1(M)$ from $p$ as a universal covering map ($\pi_1(M)$ can
be regarded as a subgroup of $\pi_1(Q)$ via $\pi_* \circ i_*$,
since $\pi_*$ is injective due to general covering map
properties, and $i_*$ is injective due to Theorem 4.3).  That
leaves $\pi \circ i$ as the remaining portion of the
principal covering map $p$, so $\pi \circ i$ is also a covering
map with fibre $\pi_1(Q) / \pi_1(M)$, i.e., $\pi_1(V) /
i_*(\pi_1(M))$.
\qed\enddemo

\enddocument